# Does the Cross-Talk Between Nonlinear Modes Limit the Performance of NFDM Systems?

Vahid Aref, Son T. Le, and Henning Buelow

Nokia Bell Labs, Lorenzstr. 10, 70435 Stuttgart, Germany, firstname.lastname@nokia-bell-labs.com

**Abstract** *We show a non-negligible cross-talk between nonlinear modes in Nonlinear Frequency-Division Multiplexed system when data is modulated over the nonlinear Fourier spectrum and transmitted over a lumped amplified fiber link. We evaluate the performance loss if the cross-talks are neglected.*

**Introduction**

Nonlinear Frequency-Division Multiplexing (NFDM) has been proposed to attain a larger spectral efficiency (SE), or to offer a low-complexity equalization method for a long-haul optical fiber system by exploiting the Kerr nonlinearity in system design [1, 2]. Over an ideally lossless fiber link, the pulse propagation is modeled, up to the first order, by the nonlinear Schrödinger equation (NLSE). Based on the well-known inverse scattering method [3], a solution of the "standard" NLSE can be decomposed into some *uncorrelated* nonlinear modes. Then, its temporal evolution maps into a *simple linear* evolution of its nonlinear modes. These two properties suggest to modulate data on some nonlinear modes, and recover them independently.

The nonlinear modes are partitioned into two parts: the continuous spectrum including the real-valued frequencies and the discrete spectrum including a set of complex nonlinear modes, called eigenvalues, characterizing the solitonic part of a pulse. The NFDM concept is experimentally verified in various scenarios [4–11]; e.g. showing the net rate of 125Gb/s and the SE of 2.3 bits/s/Hz on single polarization over 976 km link [10].

To increase the SE, we demonstrated in [11] the modulation on both spectra as, in principle, both spectra are uncorrelated. In the presence of ASE noise of amplifiers, however, the nonlinear modes become correlated. Existence of such correlation is discussed in [8, 11–13] for either of spectra in a particular scenario. Here, we study the correlation between nonlinear modes in both spectra.

In this paper, we modulate both spectra in a more general way with 55.3 Gb/s gross rate. On the continuous spectrum, we modulate 64x0.5 Gbaud OFDM signals with 32-QAM format, pre-compensated to increase the bit-rate. Each symbol has also 4 eigenvalues with the same imaginary part but different real part. Each eigenvalue is 8-PSK modulated. We experimentally demonstrate the transmission over 18x81.3 km in a SMF fiber loop with EDFAs. Using simulation, we evaluate the correlation between nonlinear modes in presence and absence of eigenvalues. We show that an eigenvalue interferes with the neighboring frequencies of the continuous spectrum. Using mutual information, We estimate the loss in the achievable rate if the correlation is neglected.

**Modulation of Nonlinear Spectrum**

The nonlinear spectrum of a pulse $q(t)$ has two parts: The continuous part, denoted by $q_c(\omega)$, is the spectral amplitude for frequencies $\omega \in \mathbb{R}$. The discrete part contains a finite set of complex eigenvalues $\{\lambda_k\} \subset \mathbb{C}^+$ and the corresponding spectral amplitudes $q_d(\lambda_k)$, defined based on Zakharov-Shabat system, (see [1]). As depicted in Fig. 1(a), we modulate data over both spectra as follows:

*A. Continuous Spectrum:* We modulate 320 bits over $N = 64$ overlapping orthogonal sub-channels (nonlinear modes) in the continuous spectrum in a similar way as the conventional OFDM signal is designed in the linear Fourier spectrum. For each symbol, the continuous spectrum is defined as:

$$q_c(\omega) = Ae^{-2j\omega^2 L} \sum_{k=-N/2}^{N/2} C_k \mathrm{sinc}(\tfrac{T_c}{\pi}\omega + k), \quad (1)$$

where $L$ is the transmission distance, $C_k$ is drawn from 32-QAM constellation, $A$ is the power control parameter, $T_c = 2$ ns is the useful block duration defining the baud-rate of 0.5Gbaud. To avoid ISI during propagation, we need a large enough guard interval which can be minimized using the pre-compensation term $e^{-2j\omega^2 L}$ [10]. Including guard intervals, the total symbol duration is $T_s = 6$ ns resulting the gross rate of 53.3 Gb/s.

*B. Discrete Spectrum:* In each 6-ns symbol, we also modulate further 12 bits over 4 eigenvalues,

$$\lambda_1 = -3\pi f_0 + j\sigma, \lambda_2 = -\pi f_0 + j\sigma$$
$$\lambda_4 = +3\pi f_0 + j\sigma, \lambda_3 = +\pi f_0 + j\sigma$$

where $\sigma = \frac{3}{2T_0}$ and $f_0 = \frac{10}{\pi T_0} \approx 3.2$ GHz after scaling time by $T_0 = 1$ ns to the physical units. The spectral amplitude of each eigenvalue is independently 8-PSK modulated, i.e. $q_d(\lambda_k) = R_k \exp(j\frac{2\pi}{8}m_k)$ with $m_k \in \{0, 1, \ldots, 7\}$.

Since the eigenvalues have non-zero real parts, the solitonic part is dispersive. For simplicity, we can imagine that each eigenvalue corresponds to a "sech" shape pulse and these pulses are non-linearly superposed in time-domain. During the transmission, the "sech" component of $\lambda_k$ walks off proportional to Re($\lambda_k$). Therefore, we should locate each "sech" pulse such that it stays (with long enough tails) in the symbol period $T_s = 6$ ns. Let $t_k$ denote the center position of the "sech" component corresponding to $\lambda_k$. Controlling $t_k$, $R_k = |q_d(\lambda_k)|$ is chosen such that $t_1 = 2.5$ ns, $t_2 = 4$ ns, $t_3 = 2$ ns and $t_4 = 3.5$ ns. By these choices, the 2 "sech" components of $\lambda_2$ and $\lambda_3$ move slowly away from center of the pulse. The other two moves first toward the center, collide at the center, and then move away from the center.

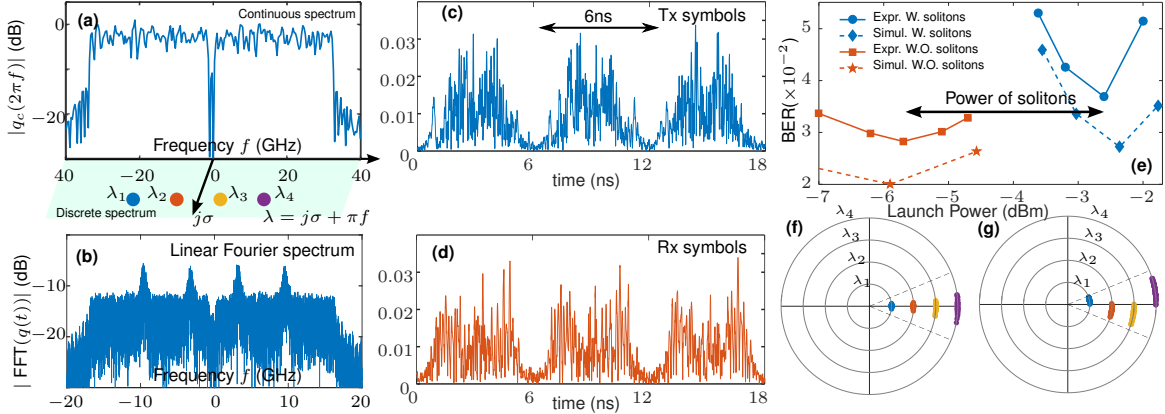

**Fig. 1:** **(a)** The nonlinear spectrum with 4 eigenvalues **(b)** The Fourier spectrum of a corresponding pulse; **(c)** A train of 6ns symbols in transmitter, and **(d)** in receiver. **(e)** The BER performance in the experiment and simulation, with/without multiplexing solitons. The error in the received phase of 4 solitons after removing 8-PSK modulation **(f)** in simulation and **(g)** in the experiment.

The symbols are generated by multiplexing the discrete spectrum, i.e. $\{\lambda_k, q_d(\lambda_k)\}_{k=1}^{4}$ and the continuous spectrum $q_c(\omega)$ using the Darboux transformation as as described in [11]. Fig. 1(b) illustrates the frequency response of generated symbols. We can clearly observe the 4-solitonic components with carrier frequencies $\pm f_0$ and $\pm 3f_0$. Figures 1(c) and 1(d) show the same train of 6ns-symbols in transmitter and receiver with no significant ISI between adjacent symbols.

**Experimental and Simulation results**

The experimental setup with a re-circulating loop is illustrated in Fig. 2. The time domain signal $q(t)$ was generated from (linear) multiplexing symbol by symbol random streams of continuous and discrete spectrum having the total power, $P = P_d + P_c$ where $P_d$ is the power of discrete spectrum and $P_c$ is the power of continuous spectrum. Determined by the eigenvalues, $P_d \approx 0.065$ mW is constant but $P_c$ is controlled by factor $A$ in (1). After resampling to 88 GSa/s, $q(t)$ is normalized according to the path-averaged model of lumped amplified links.

The fiber loop consists of 3x81.3 km spans of standard single mode fiber and EDFAs. At the receiver, after coherent detection, digital sampling at 80 GSa/s, we applied a data-aided phase and frequency offset compensation. Then, the received signal was normalized to the initial power $P$. For each symbol, we next recover the nonlinear spectrum as follows:

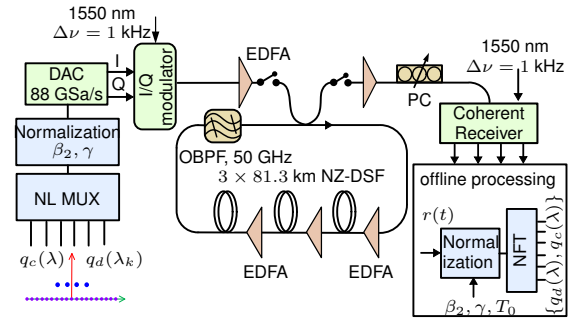

**Fig. 2:** Experimental setup with offline NFT-based detection

*A. Continuous Spectrum:* The detection of 64 sub-channels was carried out according to [9]: After recovering the received continuous spectrum by a layer-peeling-type algorithm, the single-tap phase-shift removal operation is done to remove the interplay of dispersion and nonlinearity as described in [9]. Next, the detected $q_c(\omega)$ is first mapped to the closest 32-QAM point in each sub-channel and then, mapped to the corresponding 5-bits according to an optimized 32-QAM bit-mapping.

Fig. 1(e) shows the transmission performance of 1460 km link in terms of bit-error-rate (BER). The BER curves are evaluated in two transmission scenarios: without modulation on discrete spectrum, and with 4 eigenvalues modulated. In all curves, the launch power is tuned by changing only $P_c$.

Multiplexing with discrete spectrum, the optimal power (resulting the minimum BER) is increased about 3.2 dB almost equal to the additional power $P_d$. In fact, both minimum BER is attained at the same $P_c$. Although the nonlinear modes are uncorrelated in the noiseless model, adding solitons degrades the performance of sub-channels. This indicates a non-negligible cross-talk between modes.

To study the cross-talk, we emulate the transmis-sion using the split-step Fourier simulation. The resulted BER curves are consistent with the experimental results in Fig. 1(e). Let $\hat{q}_k = \hat{y}_{2k-1} + j\hat{y}_{2k}$ denote the detected $q_c(\omega)$ at sub-channel $k$. The covariance matrix of received $\{\hat{y}_k\}_{k=1}^{128}$ illustrates the correlation between sub-channels. Figures 3(b) and 3(c) illustrate visually the covariance matrix at the optimal power in the presence and absence of solitons. Each entry is computed by

$$r_{km} = \mathbb{E}\left[(\hat{y}_k - \mathbb{E}[\hat{y}_k])(\hat{y}_m - \mathbb{E}[\hat{y}_m])\right]$$

where the average is taken over 300 different symbols, each transmitted 3000 times. The non-zero off-diagonal entries in both matrices show the correlation with $\pm 4$ adjacent sub-channels in the presence of ASE noise. By multiplexing solitons, all $r_{km}$ are increased, specially the sub-channels close to $\pm \pi f_0$ and $\pm 3\pi f_0$. This indicates that the solitons interfere into the continuous spectrum and in particular, with the closer frequencies.

Fig. 3(a) shows the achievable rate of each sub-channel assuming that the noise has the Gaussian distribution with the above covariance matrix and each sub-channel is decoded individually. We observe that in the presence of solitons, the achievable rate of all sub-channels decreases, specially the ones having the frequencies $\pm \pi f_0$ and $\pm 3\pi f_0$.

*B. Discrete Spectrum:* $q_d(\lambda_k)$ is usually detected by an NFT algorithm [5]. If two solitonic components (two "sech" shaped) are well separated in time-domain, or the corresponding eigenvalues have a relatively large difference in real part, then we can even estimate $q_d$ in time-domain by linear filtering. In fact, the received pulse is matched filtered by the soliton with the design eigenvalue to retrieve the phase. The results of this method is sometimes found more promising than the ones of applying a numerical NFT algorithm [6].

The detected phases are then followed by the single-tap phase-shift removal operation performed according to designed eigenvalues. Figures 1(f) and 1(g) show the phase errors of 4 eigenvalues after removing 8-PSK modulation. The absolute-value of the rings are chosen arbitrarily. The variance of phase-errors are small for all eigenvalues resulting a small BER<1e-4 for each phase. Interestingly, we observed an offset between the phase-errors of different eigenvalues in the experiments while no offset was observed in the simulations.

## Conclusion

We demonstrated the full modulation of nonlinear spectrum with 55.6 Gb/s gross rate over 1460 km SMF in a loop experiment. The continuous spectrum is modulated by 64x0.5 Gbaud OFDM signals with 32-QAM format, and the discrete spectrum

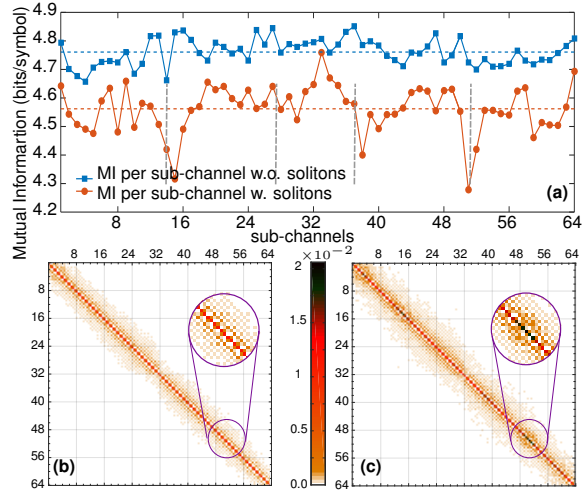

**Fig. 3:** **(a)** The achievable rate per sub-channel in the continuous spectrum when there are solitons modulated or not. The vertical dashed lines indicate $\omega = \pm \pi f_0$ and $\pm 3\pi f_0$. The coverience matrix of the received $\hat{y}_k$ of sub-channels: **(b)** when there is no soliton **(c)** when there are 4 eigenvalues modulated.

contains 4 eigenvalues, each 8-PSK modulated.

We numerically characterized the cross-talks between nonlinear modes and showed that independently decoding of nonlinear modes causes a significant performance loss, particularly, for the modes closer to the eigenvalues. An important open question is if the interference can be equalized or it is fully scrambled in the channel noise.

**References**

[1] M. I. Yousefi and F. R. Kschischang, "Information transmission using the nonlinear fourier transform, part i-iii," IEEE Trans. on Inf. Th., vol. 60, no. 7, 2014.

[2] J. E. Prilepsky, et al. "Nonlinear inverse synthesis and eigenvalue division multiplexing in optical fiber channels," Phys. Rev. Lett., vol. 113, no. 1, p. 013901, 2014.

[3] M. J. Ablowitz, et. al. "The inverse scattering transform-fourier analysis for nonlinear problems," Studies in Applied Mathematics, 1974.

[4] Z. Dong, et. al. "Nonlinear frequency division multiplexed transmissions based on nft," IEEE PTL, vol. 27, 2015.

[5] V. Aref, et. al. "Experimental demonstration of nonlinear frequency division multiplexed transmission," in 41st ECOC, 2015.

[6] H. Buelow, V. Aref, and W. Idler, "Transmission of waveforms determined by 7 eigenvalues with psk-modulated spectral amplitudes," in 42nd ECOC, Sept. 2016.

[7] A. Geisler and C. Schaeffer, "Experimental nonlinear frequency division multiplexed transmission using eigenvalues with symmetric real part," in 42nd ECOC, Sept. 2016.

[8] T. Gui, et. al. "Alternative decoding methods for optical communications based on nft," J. Lightw. Technol., 2017.

[9] S. T. Le, H. Buelow, and V. Aref, "Demonstration of 64x0.5gbaud nonlinear frequency division multiplexed transmission with 32qam," in OFC, W3J.1, 2017.

[10] S. T. Le, et. al. "125 gbps pre-compensated nonlinear frequency-division multiplexed transmission," *submitted to* ECOC, 2017.

[11] V. Aref, S. T. Le, and H. Buelow, "Demonstration of fully nonlinear spectrum modulated system in the highly nonlinear optical transmission regime," in 42nd ECOC, 2016.

[12] H. Buelow, et. al. "Experimental nonlinear frequency domain equalization of qpsk modulated 2-eigenvalues soliton," in OFC, Tu2A.3, 2016.

[13] S. Wahls, "Second order statistics of the scattering vector defining the dt nonlinear fourier transform," ITG SCC 2017.